# Direct observation of the lattice dynamics of transition metals using ultrafast electron diffraction

A. Nakamura[1], T. Shimojima[1], M. Nakano[1], Y. Iwasa[1], K. Ishizaka[1]

[1]*Quantum-Phase Electronics Center and Department of Applied Physics, the University of Tokyo*, Bunkyo, Tokyo 113-8656, Japan

We report the lattice dynamics of transition metal thin films by using the ultrafast electron diffraction. We observe a suppression of the diffraction intensity in a few picosecond after the photoexcitation, which is directly interpreted as the lattice heating via the electron-phonon interaction. The electron-phonon coupling constants for Au, Cu and Mo are quantitatively evaluated by employing the two-temperature model, which are consistent with those obtained by optical pump-probe methods. The variation in the lattice dynamics of the transition metals are systematically explained by the strength of the electron-phonon coupling, arising from the elemental dependence of the electronic structure and atomic mass.

Investigation of the nonequilibrium states in solids induced by the photoexcitation has been one of the most challenging problems in condensed matter physics. A common approach to describe the electron and lattice dynamics is the so-called two temperature model (TTM), which assumes that electrons and phonons are in thermal quasi-equilibriums with two different temperatures. When a solid is irradiated by a femtosecond laser pulse, the electrons are immediately (non-thermally) excited and quickly transferred into deeper parts of the sample. [1] The electron subsystem is then quasi-thermalized via electron-electron interactions and accordingly starts to follow the Fermi Dirac statistics characterized by the electron temperature ($T_e$). The excess energy in the electron subsystem is redistributed to the lattice through the electron-phonon interactions usually within a few picosecond, thus raising the lattice temperature ($T_l$). [2, 3]

TTM was originally proposed by Anisimov *et al*. [4] The time-dependent relation between $T_e$ and $T_l$ can be written in the form

$$\gamma T_e \frac{\partial T_e(t)}{\partial t} = -g(T_e - T_l) + P(t), \quad (1a)$$

$$C_l \frac{\partial T_l(t)}{\partial t} = g(T_e - T_l), \quad (1b)$$

where $\gamma$, $C_l$, and $P(t)$ are the electronic specific heat coefficient, the lattice heat capacity, and the absorbed laser power, respectively, per unit volume. The electron-phonon coupling constant $g$ is often expressed in an approximate form [5, 6]:

$$g \propto N(0)\langle I^2 \rangle / M, \quad (2)$$

where $N(0)$ is the electronic density of states at the Fermi level, $M$ is the atomic mass, and $\langle I^2 \rangle$ is the Fermi surface averaged electron–phonon matrix element. The importance of the electron-phonon coupling in metals has been often discussed in terms of the polaronic effect that increases the effective mass of carriers, and also the superconductivity as raised in the theories of McMillan and Eliashberg. [6]

Pump-probe methods employing the femtosecond pulse laser provide fundamental information on the electron-phonon interactions. [5] Over the last couple of decades, transient optical reflectivity measurements using TTM analysis have significantly contributed to the estimation of $g$ values in thin-film metals, including superconducting materials. [2,7] However, the quantitative estimation of $T_e$ from the transient optical reflectivity data has been of difficulty especially for transition metals, due to the complex electronic structure including both *s/p* and *d* orbitals near the Fermi level. The evaluation of $g$ by using various kinds of probes beyond the optical methods, especially the ones including the defection of $T_l$, has been long desired.

Dynamical diffraction measurements can detect $T_l$ through the Debye-Waller effect, by recording the transient diffraction intensity. Recently, the ultrafast electron diffraction (UED) technique [8, 9] has been developed for investigating the transient lattice dynamics of metal thin films. In the weak excitation regime, $g$ values of noble metals such as Au and Ag has been actually estimated from the UED data by applying TTM analysis [10]. On the other hand, under the extremely strong laser excitation, the mechanism of the ultrafast melting of Au and Al nano-films has been also investigated [11, 12]. While it has been shown that UED is a suitable technique for evaluating the lattice dynamics, systematic element dependence of electron-phonon coupling constant $g$ for the various transition metals has not been investigated.

In this paper, we report the lattice dynamics of thin films of Au, Cu and Mo metals by using UED. Electron-phonon coupling constants for these transition metals are quantitatively evaluated by employing the TTM analysis. Variation in the lattice dynamics in the transition metals are systematically explained by the strength of the electron-

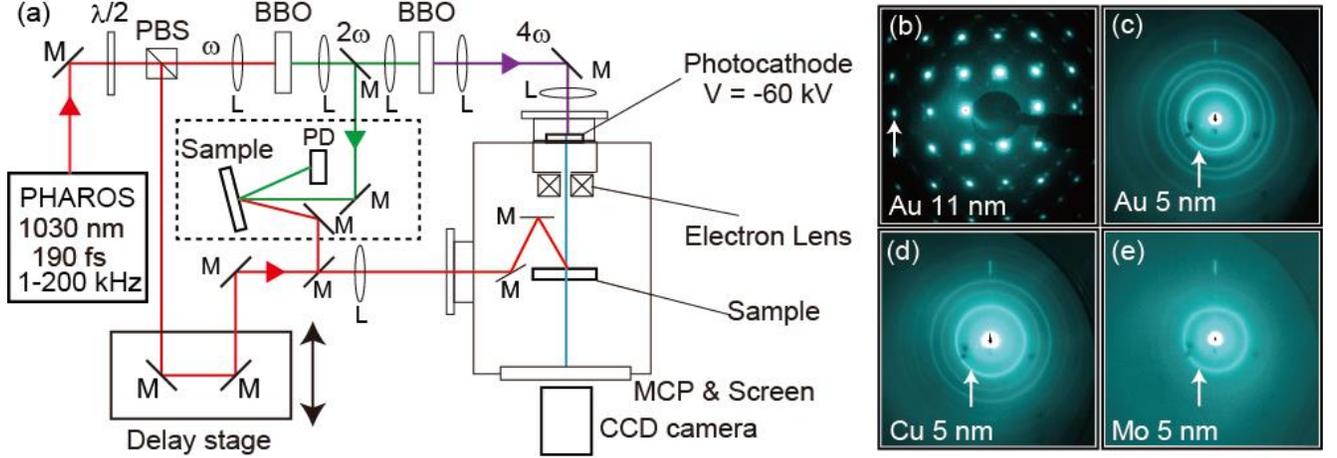

FIG. 1 (a) Schematic experimental setups for UED and transient optical reflectivity measurements. The fundamental laser pulse (1.2 eV) is split into two branches. One is passed through a delay line for exciting the sample. The other part is utilized to generate the frequency quadrupled ultraviolet pulse (4.8eV) for driving the electron gun. Electron pulses are emitted from a photocathode and accelerated to 60 keV. Electron diffraction patterns are obtained using a phosphor screen, MCP and CCD camera. (b-e) The static electron diffraction patterns for a single crystalline Au, polycrystalline Au, Cu, and Mo, respectively. These diffraction patterns were obtained at room temperature with the repetition rate of 200 kHz. White arrows indicate the diffraction spot and rings used for the analysis in Fig.2-4.

phonon coupling, reflecting the element-dependent electronic structure and atomic mass.

A schematic of the experimental setup for UED is shown in Fig. 1(a). It consists of the femtosecond (190 fs) laser system (PHAROS, Light Conversion) and the ultrahigh-vacuum ($\sim 10^{-10}$ Torr) chamber for the diffraction measurement. The generated laser is split into two beams, pump and probe, by a polarized beam splitter (PBS). The fundamental 1030 nm photon pulse is used to excite the sample and the frequency quadrupled 257 nm photon pulse to produce the electron packets for probe. The generated electron packets are accelerated to 60 keV and focused onto the sample by a magnetic lens. In a transmission geometry, the diffracted electrons were recorded by using a phosphor screen with a microchannel plate detector (MCP) and charge-coupled device (CCD) camera. The total time resolution of the system $\Delta t_{total}$ is 650 fs.

Transient optical reflectivity measurements were performed using the same optical setup. As shown in Fig. 1(a), the frequency doubled 515 nm photon pulse is used to probe the sample which is photoexcited by the 1030 nm photon pulse. The reflectivity was measured by using a photo detector (PD).

Polycrystalline films of Au, Cu, and Mo were prepared by depositing onto the copper micro grids covered with carbon membranes for UED measurements, and onto thin glass substrates for optical reflectivity measurements. Deposition rate was set to 0.1 Å/s and the thickness of the film is 5 nm. A free standing single crystalline Au film with a thickness of 11 nm (Oken Shoji) fixed on a copper micro grid was also used for UED measurements.

We begin with the UED results obtained for the single crystalline Au with thickness of 11 nm to quantitatively determine the transient $T_l$ and the value of $g$. The static electron diffraction pattern in Fig. 1(b) was obtained at room temperature (RT) with a repetition rate of 200 kHz. The intense Bragg peaks indicate the reciprocal lattice of the face-centered cubic (fcc) structure. Figure 2(a) shows the time evolution of the relative (600)-peak intensity, $\Delta I(t)/I_0 = [I(t) - I_0] / I_0$, where $I(t)$ and $I_0$ are the integrated intensities of (600) peak at time $t$ and $t < 0$ (before the photoexcitation), respectively. They are obtained at a repetition rate of 10 kHz under the pump fluence of 0.8 mJ/cm$^2$. The (600)-peak intensity is suppressed by $\sim 5$ % within $\sim 10$ ps. In this weak excitation regime, the suppression of the Bragg peak intensity can be regarded as the signature of the lattice thermalization, and expressed by using the Debye-Waller-factor temperature parameter $Y(T_l)$ [10] as

$$\log_{10}(1 + \Delta I / I_0) = (\sin\theta / L)^2 Y(T_l), \qquad (3)$$

where $L$ is the wave length of the probe electron beam (0.491 pm) and $\theta$ is the Bragg angle. Within the high temperature limit Debye model, $Y(T_l)$ is expressed as

$$Y(T_l) = -(\log_{10} e)\frac{48\pi^2 \hbar^2}{M k_B \Theta_D^2}(T_l - T_0), \qquad (4)$$

where $\hbar$ is the reduced Planck constant, $k_B$ is the Boltzmann constant, $T_0$ is the original lattice temperature before the photoexcitation ($\Delta I = 0$), and $\Theta_D$ is the Debye temperature. Applying Eqs. (3) and (4), $\Delta I(t)/I_0$ can be directly related with the transient lattice temperature $T_l(t)$. The Debye-Waller coefficient $dY/dT_l$ in Eq. (4) is evaluated from static

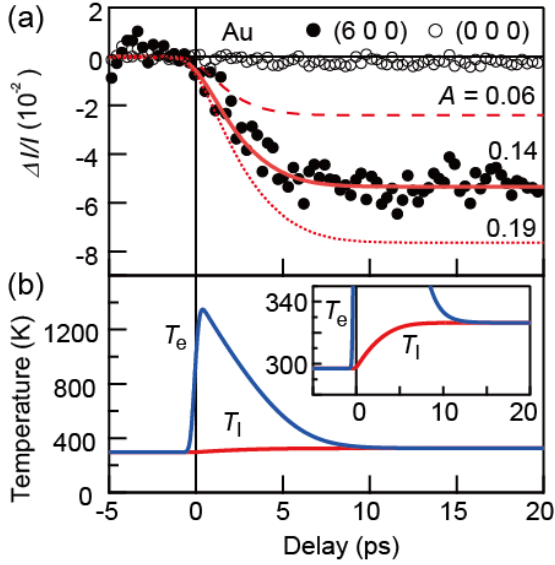

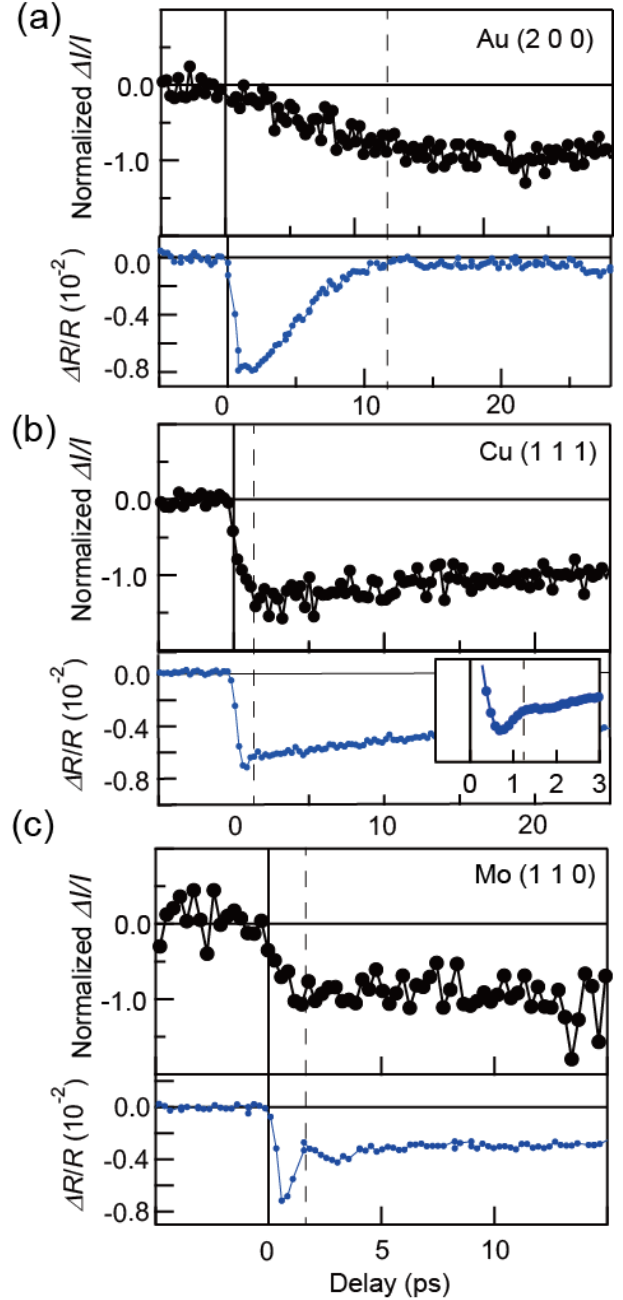

Fig. 2 (a) The relative intensity change of (600) Bragg spot [indicated by the white arrow in Fig. 1(b)] for a single crystalline Au film of 11 nm obtained at a repetition rate of 10 kHz with a fluence of 0.8 mJ/cm$^2$. Solid red curve represents the best fitting result by the TTM analysis, where the evaluated values of $g$ and $A$ are $1.8 \pm 0.5$ ($10^{16}$ W/m$^3$K) and $0.14 \pm 0.02$, respectively. Those with $A = 0.06$ and $0.19$ are also displayed as the references shown by the broken curves. (b) The calculated $T_e$ and $T_l$ with $g = 1.8$ ($10^{16}$ W/m$^3$K) and $A = 0.14$. Other fixed parameters are listed in Table 1. Both $T_e$ and $T_l$ reach the quasi-equilibrium temperature 326 K at ~10 ps after photoexcitation.

x-ray diffraction measurements at several temperatures and theoretical calculations. [13, 14] The values of $dY/dT_l$ we used for Au is $-1.9 \times 10^{-3}$ Å$^2$/K, which corresponds to $\Theta_D = 157$ K.

For the estimation of $g$, $\Delta I(t)/I_0$ curves were fitted by using the $T_l(t)$ solutions of the TTM [Eq. (1)] convolved with a Gaussian including the total time resolution of the system ($\Delta t_{total} = 650$ fs). The source term $P(t)$ representing the volume-averaged absorbed pump laser power in Eq. (1) is given by

$$P(t) = AF(t)/d, \quad (5)$$

where $d$ is the film thickness and $F(t)$ is the pump laser flux per unit area which is a Gaussian function of $t$ with the width of $\Delta t_{pump} = 190$ fs. $A$ is expressed as $A = 1 - R - T$, where $R$ and $T$ is the reflectivity and transmissivity of the thin-film sample, respectively. Since the film thickness of 11 nm is much shorter than the ballistic range of the hot electrons in typical metals (20-100 nm) [15], we do not need to consider the dissipation process of the hot electrons into the deeper part of the sample. We note that the multiple scattering effect is negligible in our experimental condition [16]. It has been also pointed out that the electronic heat

Fig. 3 (a-c) UED and the transient optical reflectivity data for polycrystalline Au, Cu and Mo films of 5 nm, respectively. $\Delta I/I$ is normalized by that at 20 ps for each material. These data were obtained at a repetition rate of 10 kHz with a fluence of 0.95 mJ/cm$^2$. Diffraction rings used for the analysis are indicated by the white arrows in Fig. 1(c-e). The broken lines represent the timescale of the lattice heating.

capacity is a non-monotonic function of $T_e$, depending on the density of states of the material [11,17,18]. According to ref. 17, however, our experimental condition which raises $T_e$ to ~1300 K at the highest allows us to use the constant $\gamma$ values of RT in Eq. (1a). Similarly, we assume the $t$-independent $g$ and $C_l$ values in this work [17].

For the analysis, we set $g$ and $A$ as the fitting parameters, whereas $dY/dT_l$ and $\gamma$ are fixed as given in Table 1. The $C_l$ values are calculated by using Dulong-Petit law for all samples ($C_l = 25$ JK$^{-1}$mol$^{-1}$). We note that one can obtain a unique pair of $g$ and $A$ values since the former is independently related with the time-scale of the $\Delta I/I$ curve while the latter only affects the height of $\Delta I/I$. This tendency is evidenced in the three curves with a fixed $g$ value [$g = 1.8 \pm 0.5$ (10$^{16}$ W/m$^3$K)] and three different $A$ values (0.06, 0.14, and 0.19) in Fig. 2(a). The solid red curve in Fig. 2(a) represents the best fitting result with $A = 0.14 \pm 0.02$. The value of $A$ should naturally depend on the detailed form (especially the thickness) of the thin-film sample, but is consistent with the results reported so far. [19] $T_e$ and $T_l$ thus obtained by the TTM analysis are displayed in Fig. 2(b). $T_l$ increases monotonically from RT to the quasi-equilibrium temperature ($T_q$) of 326 ± 5 K. On the other hand, after the rapid increase of $T_e$ to ~1300 K immediately after the photoexcitation, $T_e$ decreases down to $T_q$ in ~10 ps. These are consistent with the picture presented in the previous report. [15]

In order to investigate the element dependence of the electron-phonon coupling, we applied the TTM analysis to the UED data for several transition metals. Figures 1(c-e) show the static electron diffraction images for polycrystalline 5 nm thin films of Au, Cu, and Mo, respectively. The diffraction rings were identified to be fcc structure for Au and Cu, and body-centered cubic (bcc) structure for Mo. $\Delta I/I$ of particular indices, obtained by integrating the intensity of each ring, are presented by the black markers in the upper panels of Fig. 3(a-c). $\Delta I/I$ curves exhibit the rapid suppression of Bragg intensities after the photoexcitation with different time scales, depending on the element.

By looking at the UED and transient optical reflectivity data as shown in upper and lower panels in Fig. 3, respectively, we can discuss the $t$-evolution of $T_e$ and $T_l$. The transient optical reflectivity, being sensitive to the change in $T_e$, exhibits a relaxation time that is comparable to the rise time of $\Delta I/I$, as indicated by the dotted lines in Fig. 3(a-c). These provide the time-scale where the electron and lattice subsystems attain the common quasi-equilibrium state ($T_e \sim T_l \sim T_q$). It is clearly shown that the lattice heating due to the energy transfer from the electron to the lattice subsystems occurs in Au more slowly than that in Cu and Mo, suggestive of the weaker electron-phonon coupling for Au.

To obtain the $g$ values, $\Delta I/I$ curves were analyzed using the TTM as in the case for the single crystalline Au. The fixed parameters are summarized in Table 1. As shown in Fig. 4(a-c), $\Delta I/I$ curves for Au, Cu and Mo were well reproduced by the fitting curves represented by the solid red curves. The obtained $g$ value for Au, $1.5 \pm 1 \times 10^{16}$ W/m$^3$K, agrees well with the result on the 11 nm single-

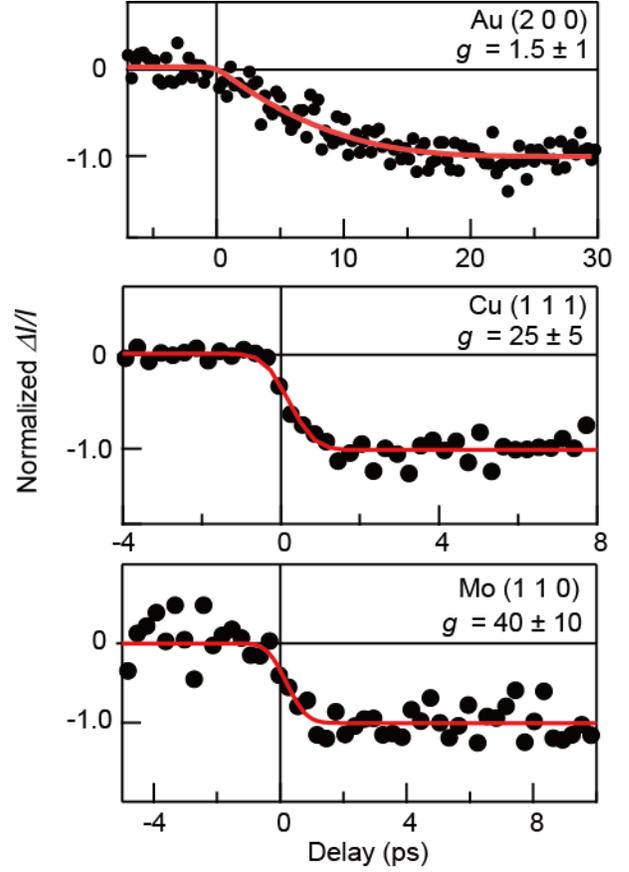

Fig. 4 (a-c) UED data with fitting curves for polycrystalline Au, Cu, and Mo, respectively. Red solid curves represent the best fit obtained by the TTM analysis with the $g$ values of 1.5, 25 and 40 × 10$^{16}$ W/m$^3$K, respectively.

crystalline sample, and is found to be much smaller than those for Cu (25 ± 5 × 10$^{16}$ W/m$^3$K) and Mo (40 ± 10 × 10$^{16}$ W/m$^3$K). We note that the $g$ values for these metals calculated by using the linear response method give a similar tendency. [20] The $g$ values for various cubic metals have been also previously evaluated by using the transient optical reflectivity measurements. [2, 7, 15, 21] Regarding Au, there are several studies [see Ref. 15 and references therein] reporting the $g$ values of ~2 × 10$^{16}$ W/m$^3$K, which are quantitatively consistent with the present UED result. The $g$ values reported for Cu (10 × 10$^{16}$ W/m$^3$K) [2] and Mo (13 × 10$^{16}$ W/m$^3$K) [21] are again much larger than that of Au, giving the qualitative similarity with UED. Quantitatively, however, they are somewhat smaller compared to the present work. It may be partly due to the difficulty of the $T_e$ estimation from the optical reflectivity in the strong electron-phonon coupled system. The present UED result will offer a new aspect toward the more quantitative understanding of electron-phonon interaction.

| | $\gamma$ (J/m$^3$K$^2$) [24] | d$Y$/d$T_l$ (10$^{-3}$Å$^2$/K$^2$) [13, 14, 23] | $g$ (10$^{16}$W/m$^3$K) Present work |
|---|---|---|---|
| Au | 71 | -1.9 | 1.5 ± 0.5 (1.8 ± 0.5 for single crystal) |
| Cu | 98 | -2.0 | 25 ± 5 |
| Mo | 211 | -0.5 | 40 ± 10 |

Table 1. The values of the electronic specific heat coefficient $\gamma$ and the Debye-Waller factor coefficient d$Y$/d$T_l$ used for the TTM fitting analysis, and the electron-phonon coupling constant $g$ obtained in the present work. $\gamma$ and d$Y$/d$T_l$ are taken from Ref 13, 14, 23, 24.

Finally, we discuss the element dependence of the $g$ values obtained by UED. Equation (2) shows that $g$ depends on the electronic density of states $N(0)$, the Fermi-surface-averaged electron–phonon matrix element $<I^2>$, and the atomic mass $M$. The large value of $g$ for Mo is explained by the large $N(0)$ compared to those of Au and Cu due to the partial density of states of the Mo 3$d$ orbital on the Fermi level. This tendency is also seen in the $\gamma$ values in Table 1. On the other hand, the difference in the $g$ values for Au and Cu should be attributed to $M$, considering that the noble metals have comparable $<I^2>$ and $N(0)$ values. [20] This tells us that the so-called Hopfield parameter [22] $\eta = N(0)<I^2>$ is useful to understand the large $g$ value for $d$ electron system while the effective spring constant $M<\omega^2>$ could be important when comparing the $g$ values for $s/p$ electron systems.

In conclusion, we performed the UED and transient reflectivity measurement on Au, Cu, and Mo thin films. The present work shows that the transient lattice temperature and electron-phonon coupling constant can be determined by using the UED method and the TTM. The time-scale of the lattice dynamics in transition metals are systematically explained by the difference in the $g$ values arising from the element-dependent electronic structure and atomic mass. The combination of TTM and UED will be useful for the investigation of various materials where the electron-phonon coupling plays an important role, such as superconductors, high-mobility semiconductors, systems with Peierls instability, and so on.